\documentclass[aps,prl,twocolumn,showpacs,superscriptaddress,groupedaddress]{revtex4-2}  
\usepackage{graphicx}  
\usepackage{dcolumn}   
\usepackage{bm}        
\usepackage{amssymb}   
\usepackage{amsmath}   
\usepackage{braket}    
\usepackage{mathtools}
\usepackage[hidelinks,linktocpage]{hyperref} 
\usepackage{cleveref}
\usepackage{subfig}
\usepackage{float}
\usepackage{xcolor}

\usepackage{fancyhdr}
\begin{document}

\fancyhead[R]{\ifnum\value{page}<2\relax\else\thepage\fi}

\title{Unconditional remote entanglement using second-harmonic generation \\
and twin two-mode squeezed vacuum states}
\author{Richard J. Birrittella$^{1,2}$, James Schneeloch$^{1}$, Christopher C. Tison$^{1}$, Michael L. Fanto$^{1}$, Paul M. Alsing$^{1}$ and Christopher C. Gerry$^{3}$ \\
\textit{\textit{$^{1}$Air Force Research Laboratory, Information Directorate, Rome, NY, USA, 13441}
	\\
\textit{$^{2}$National Academy of Science, 500 Fifth St. N. W., Washington DC 20001, USA\\}	
\textit{$^{3}$Department of Physics and Astronomy, Lehman College,\\
The City University of New York, Bronx, New York, 10468-1589,USA} 
}}

\date{\today}

\begin{abstract}
We propose a photonics-based, continuous-variable (CV) form of remote entanglement utilizing strictly second-order nonlinear optical interactions that does not require the implementation of a state-projective measurement (i.e. remote entanglement without conditioning). This scheme makes use of two separate down-converters, wherein the corresponding nonlinear crystals are driven by strong classical fields as prescribed by the parametric approximation, as well as a fully quantum-mechanical model of nondegenerate second harmonic generation (SHG) whose evolution is described by the trilinear Hamiltonian of the form $\hat{H}_{\text{shg}} = i\hbar\kappa\big(\hat{a}\hat{b}\hat{c}^{\dagger} - \hat{a}^{\dagger}\hat{b}^{\dagger}\hat{c}\big)$.  By driving the SHG process with the signal modes of the two down-converters, we show entanglement formation between the generated second-harmonic mode (SH-mode) and the non-interacting joint-idler subsystem without the need for any state-reductive measurements on the interacting modes. 
\end{abstract}


\maketitle

\thispagestyle{fancy}
\section{\label{sec:Intro} I. Introduction}

\noindent Remote entanglement can encompass a broad range of techniques in which separate parties can be made to be entangled without ever directly interacting with one another and serves as an important tool in the development of future quantum networks where entanglement between spatially separated parties beyond the reach of direct transmission becomes especially important \cite{ref:Wehner,ref:Zeilinger,ref:Basset}. One such example of this is the original entanglement swapping experiment performed by Pan \textit{et al.} \cite{ref:Pan} using spontaneous parametric down-conversion (SPDC) sources. In their experiment they subject two pairs of polarization-entangled photons to a Bell-state measurement, projecting the remaining photons into an entangled state. Similarly, and more recently, Park \textit{et al.} \cite{ref:Park} devised an entanglement swapping scheme using two pairs of polarization-entangled photons generated via spontaneous four-wave mixing (SFWM) in a Doppler-broadened $\prescript{87}{}{\text{Rb}}$ atomic ensemble. They went on to show a violation of a Clauser-Horne-Shimony-Holt (CHSH) Bell's inequality using the entanglement-swapped pair of photons. Two-photon interference has also been exploited in entanglement-swapping schemes to entangle spins \cite{ref:Bernien}, atoms \cite{ref:Rosenfeld} and using photons generated via quantum dots \cite{ref:Basset}. 

We propose a means of forming unconditional entanglement (i.e., without heralding or post-selection) between non-interacting optical modes using a second-order nonlinear interaction. More specifically, we consider performing (spatially) non-degenerate second-harmonic generation (SHG) using the signal modes of independent two-mode squeezed vacuum states generated via SPDC. We show entanglement formation between the SH-mode and the spatially-separated non-interacting joint-idler subsystem.  This entanglement occurs without the need for any state-projective measurements, persists for experimentally accessible interaction times, and scales reasonably with input average photon number.

There has been extensive work done detailing a quantum model of SHG.  Typically one considers the degenerate case in which two photons of frequency $\omega$, occupying the same optical mode, are annihilated to produce a single photon with frequency $2\omega$.   The corresponding Hamiltonian driving this interaction is of the form $\hat{H}_{\text{shg}} = i\hbar\kappa\left(\hat{a}^{2}\hat{c}^{\dagger}-\hat{a}^{\dagger\;2}\hat{c}\right)$ which has no exact solution \footnote{Although the SHG hamiltonian cannot be completely diagonalized, individual eigenstates can be found without truncating the state space. Consider for example the Bell-state superposition between (two photons in $\hat{a}$ and none in $\hat{c}$) and (no photons in $\hat{a}$ and one in $\hat{c}$).}, nor can a direct application of the Perelomov formalism be applied \cite{ref:Perelomov}, owing to the lack of a finite-dimensional Lie algebra \cite{ref:Puri}. Still, SHG has been studied perturbatively \cite{ref:Kozierowski,ref:Chesi}, numerically \cite{ref:Ekert} resulting in the generation of sub-Poissonian light \cite{ref:Bajer,ref:Bajer2}, in terms of a factorization of photon-number moments \cite{ref:Crosignani,ref:Chmela} and within the analogous case of the Dicke model \cite{ref:Orszag}.  For our purposes, however, the interaction we consider is the (spatially) non-degenerate case in which two photons, each with frequency $\omega$, are annihilated from two different spatial modes to generate a second-harmonic photon of frequency $2\omega$.

In this model of SHG, the trilinear Hamiltonian that drives the interaction between the three field states is given by
\begin{equation}
	\hat{H}_{I} = i\hbar\kappa\big(\hat{a}\hat{b}\hat{c}^{\dagger} - \hat{a}^{\dagger}\hat{b}^{\dagger}\hat{c}\big),
	\label{eqn:trilinear}
\end{equation}

\begin{figure}
	\centering
	\includegraphics[width=1.0\linewidth,keepaspectratio]{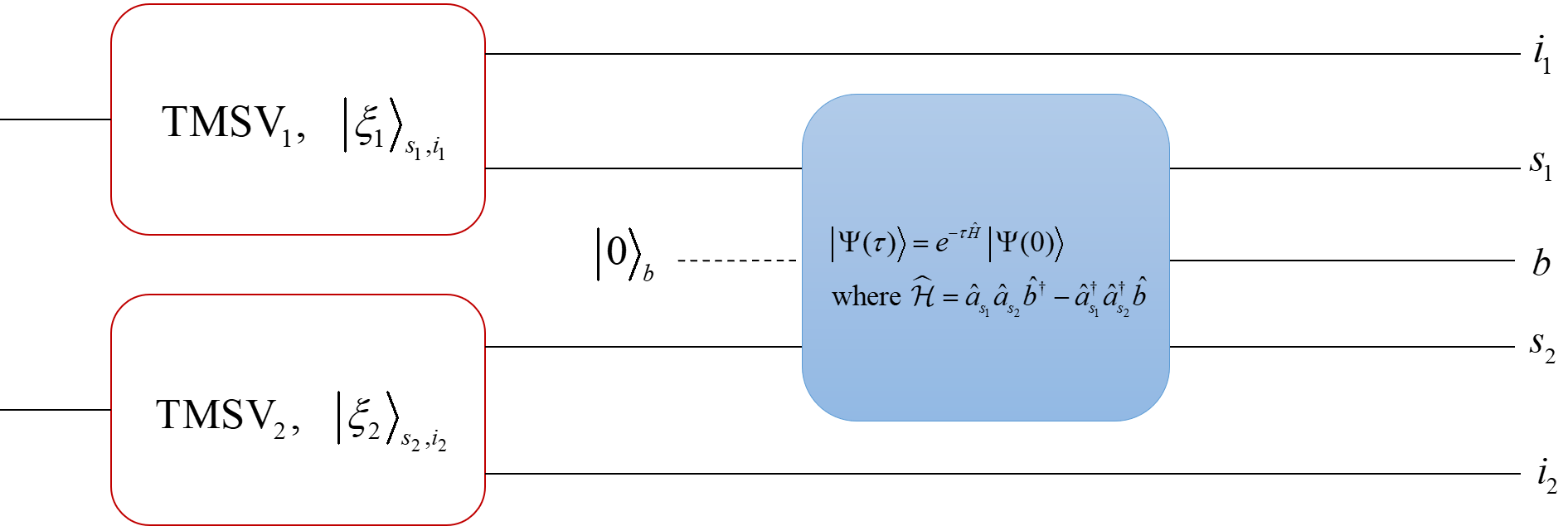}
	\caption{A schematic representation of the procedure being implemented. Two down-converters, each producing a TMSVS, are oriented such that the signal modes of each are driving the spatially non-degenerate SHG process. For our purposes, the second-harmonic $b$-mode is initially in a vacuum state.}
	\label{fig:2.1}
\end{figure}

\noindent where  $\{\hat{a},\hat{a}^{\dagger}\}$ and $\{\hat{b},\hat{b}^{\dagger}\}$ operate on the two input modes while $\{\hat{c},\hat{c}^{\dagger}\}$ operates on the SH-mode and where the parameter $\kappa$ is a coupling constant proportional to the $\chi^{\left(2\right)}$ nonlinear susceptibility \cite{ref:Boyd}. 

Although a complete quantum model characterizing interactions driven by the trilinear Hamiltonian is generally computationally intractable \cite{ref:Braunstein,ref:Niu,ref:vanLoock}, it has been investigated numerically as early as the 1960s both in the context of SPDC \cite{ref:Walls,ref:Mollow,ref:Travis,ref:Tucker,ref:Agarwal,ref:McNeil} as well as emission from super-radiant Dicke-states \cite{ref:Dicke,ref:Bonafacio}. The former, \cite{ref:Walls}$-$\cite{ref:McNeil}, investigated the eigenvalues of the tri-diagonal matrix representation of Eq. \ref{eqn:trilinear} in the computational `logical' basis $\ket{n}_{L}$ formed from the three-modes of a down-converter $\ket{n}_{L}=\ket{n_{p_{0}}}_{p}\ket{n}_{s}\ket{n}_{i}$, where $n_{p_{0}}$ are the initial number of photons occupying the pump mode.  The latter, \cite{ref:Dicke}\cite{ref:Bonafacio}, employed the Schwinger realization of the SU(2) Lie algebra \cite{ref:Yurke} to convert the pump-idler modes into the spin-boson representation such that the trilinear Hamiltonian can be written as  $\hat{H}_{I}=i\hbar\kappa\big(\hat{J}^{\left(p,i\right)}_{+}\hat{a}-\hat{J}_{-}^{\left(p,i\right)}\hat{a}\big)$.  They went on to develop differential-difference equations for the state probability amplitudes $c_{n}=_{L}\braket{n|e^{-i\hat{H}_{I}t/\hbar}|\psi_{\text{in}}}$ of the output state $\ket{\psi}_{\text{out}}=\sum_{n=0}^{\infty}c_{n}\ket{n}_{L}$. 

The trilinear Hamiltonian can also be expressed in terms of the SU(1,1) Lie Algebra, whereby the signal-idler modes are written in terms of the SU(1,1) ladder operators $K_{+}^{\left(s,i\right)}=\hat{a}^{\dagger}\hat{b}^{\dagger}$ and $\hat{K}_{-}^{\left(s,i\right)}=\hat{a}\hat{b}$ \cite{ref:Yurke} such that Eq. \ref{eqn:trilinear} becomes  $\hat{H}_{I}=i\hbar\kappa\big(\hat{c}^{\dagger}\hat{K}_{-}^{\left(s,i\right)}-\hat{c}\hat{K}_{+}^{\left(s,i\right)}\big)$. This was the form of the interaction Hamiltonian considered by Nation and Blencowe \cite{ref:Nation} and Alsing \cite{ref:Alsing} in their long-time approximation of the state statistics with a pump field taken as an arbitrary pure state. Simulations of non-degenerate parametric down-conversion were performed by Ding \textit{et al.} \cite{ref:Ding} using a linear trapped three-ion crystal simulating light-atom interactions described by the Tavis-Cummings model.  Moreover, Yanagimoto \textit{et al.} \cite{ref:yanagimoto2020broadband} explored the dynamics of broadband parametric down- conversion in the context of nonlinear nanophotonics in the regime in which the pump occupation is very small (few photons).  They went on to demonstrate Fano interference in linearly coupled $\chi^{\left(2\right)}$ waveguides.  In an effort to circumvent the intractable nature of the Hamiltonian due to the immense size of the Hilbert space, they provided a diagonlized form of the trilinear Hamiltonian in a truncated space in which single-photon down-conversion occurs.  

Second order nonlinearities have also been utilized in the context of entanglement formation. Podoshvedov \textit{et al.} \cite{ref:Podoshvedov} devised a means of heralding macroscopic entanglement from a coherent pump beam; while more recently, Birrittella \textit{et al.} \cite{ref:Birrittella} considered coherently-stimulated down-conversion with a quantized pump field and investigated the phase-dependent photon statistics of-, and entanglement properties between-, the three fields.

This paper is organized as follows: In Section \hyperref[sec:RemoteEntanglement]{II\ref*{sec:RemoteEntanglement}} we discuss the setup and provide a perturbative analysis of the time-evolved system.  We also include a qualitative analysis of the photon statistics of the second-harmonic and signal modes as well as a numerical and qualitative analysis of the entanglement properties within the system. We close the paper in Section \hyperref[sec:Conclusion]{III\ref*{sec:Conclusion}} with a brief discussion of our findings and some closing remarks.

\section{\label{sec:RemoteEntanglement} II. Formation of Unconditioned Remote Entanglement}

\subsection{\label{sec:TMSV} a.) The two-mode squeezed vacuum state}
\noindent As is well known, non-degenerate parametric down-conversion has been for years a reliable source of two-mode non-classical states of light in the laboratory \cite{ref:Drummond}.  In the parametric approximation wherein the pump field is assumed undepleted, the interaction Hamiltonian for the down-conversion process is given by \cite{ref:Drummond}\cite{ref:GerryBook}

\begin{equation}
	\hat{H}_{\text{sq}} = i\hbar\left(\gamma\hat{a}\hat{b} - \gamma^{*}\hat{a}^{\dagger}\hat{b}^{\dagger}\right).
	\label{eqn:7}
\end{equation}

The parameter $\gamma$ is proportional to the second-order nonlinear susceptibility $\chi^{\left(2\right)}$ and to the amplitude and phase of the pump laser field, assumed here to be sufficiently bright to be treated as a classical amplitude such that depletion and fluctuations in the field can be ignored. The quantized field modes $a$ and $b$ are taken to be the signal and idler fields, respectively.  The two-mode squeeze operator $\hat{S}\left(z\right)$ is realized as \cite{ref:Drummond}\cite{ref:GerryBook}

\begin{equation}
	\hat{S}\left(z\right) = e^{-i\hat{H}_{sq}t/\hbar} = e^{r\left(\hat{a}\hat{b}e^{-2i\phi}- \hat{a}^{\dagger}\hat{b}^{\dagger}e^{2i\phi}\right)},
	\label{eqn:1}
\end{equation}

\noindent where we have written $\gamma=|\gamma|e^{2i\phi}$ and where $r=|\gamma|t$ is the squeezing parameter.  Typically the signal and idler beams are initially in vacuum states, and thus the output state will be the two-mode squeezed vacuum state (TMSVS) $\ket{\xi}$ given by

\begin{align}
	\ket{\xi} &= \hat{S}\left(z\right)\ket{0}_{a}\ket{0}_{b} = \left(1-|z|^{2}\right)^{1/2}\sum_{n=0}^{\infty}z^{n}\ket{n}_{a}\ket{n}_{b}\nonumber \\
	&= \frac{1}{\cosh r}\sum_{n=0}^{\infty}\left(-1\right)^{n}e^{2in\phi}\tanh^{n}r\ket{n}_{a}\ket{n}_{b}.
	\label{eqn:4}
\end{align}

\noindent Note that $\gamma$ and $2\phi$ are the amplitude and phase of the classical pump field, respectively, and $z=e^{i2\phi}\tanh\left(r\right)$ contrained to $0\leq |z| \leq 1$.  The total average photon number is given by

\begin{widetext}
	\begin{align}
		\bar{n}_{\text{total}} &=\braket{\psi_{\text{in}}|\hat{S}^{\dagger}\left(z\right)\left(\hat{a}^{\dagger}\hat{a} + \hat{b}^{\dagger}\hat{b}\right)\hat{S}\left(z\right)|\psi_{\text{in}}} \nonumber \\
		&= \braket{\psi_{\text{in}}|\bigg[\left(\hat{a}^{\dagger}\hat{a} + \hat{b}^{\dagger}\hat{b}\right)\cosh 2r -  \left(e^{2i\phi}\hat{a}^{\dagger}\hat{b}^{\dagger}+e^{-2i\phi}\hat{a}\hat{b}\right)\sinh 2r + 2\sinh^{2}r\bigg]|\psi_{\text{in}}}
		\label{eqn:2}
	\end{align}
\end{widetext}

\noindent where we have used the operator relations

\begin{equation}
	\hat{S}^{\dagger}\left(z\right)
	\begin{pmatrix}
		\hat{a}  \\
		\hat{b} 
	\end{pmatrix} 
	\hat{S}\left(z\right)
	=
	\begin{pmatrix}
		\hat{a}\cosh r - e^{2i\phi}\hat{b}^{\dagger}\sinh r  \\
		\hat{b}\cosh r - e^{2i\phi}\hat{a}^{\dagger}\sinh r 
	\end{pmatrix},
	\label{eqn:3} 
\end{equation}

\noindent obtained by use of the Baker-Hausdorff lemma \cite{ref:Achilles}.  For the case of an input double vacuum state $\ket{\psi_{\text{in}}} = \ket{0,0}_{a,b}$ the total average photon number is that of the TMSVS, given by $\bar{n}=2\sinh^{2}r$, which is notably independent of the pump phase. The Fock states of each mode are tightly correlated and the state as a whole is highly non-classical due to the presence of squeezing in one of the two-mode quadrature operators. The joint-photon number probability distribution for finding $n_{1}$ photons in the $a$-mode and $n_{2}$ photons in the $b$-mode is

\begin{equation}
	P\left(n_{1},n_{2}\right) = \big|\braket{n_{1},n_{2}|\xi}\big|^{2} = \frac{\tanh^{2n_1}r}{\cosh^{2}r}\times\delta_{n_{1},n_{2}},
	\label{eqn:5}
\end{equation}

\noindent such that only the diagonal elements satisfying $n_{1}=n_{2}$ are nonzero.  The photon-number statistics are super-Poissonian in each mode; tracing over either mode yields a single-mode mixed state with a thermal distribution \cite{ref:Barnett} \cite{ref:Yurke2}. 

\subsection{\label{sec:perturbation} b.) Short-time approximation of the state, post-SHG}

\noindent A schematic of the proposed setup can be seen in Fig.~\ref{fig:2.1}.  We start with a pair of TMSVS, $\ket{\xi_j}$, $j=1(2)$, of equal average photon numbers such that the signal modes of each are given by $\bar{n}_{s_1}=\bar{n}_{s_2}\equiv\bar{n}_s$. The signal modes are then used to seed SHG (with the SH-mode initially in a vacuum state).  We develop the theory assuming an initial state given by 

\begin{align}
	\ket{\Psi\!\left(0\right)} &= \ket{\xi_{1}}_{s_1,i_1}\otimes\ket{\xi_2}_{s_2,i_2}\otimes\ket{\phi}_b \label{eqn:2.1} \\ &\!\!\!\!\!=\sum_{n=0}^{\infty}\sum_{n'=0}^{\infty}\sum_{m=0}^{\infty}\!\!C_{n,n',m}^{\left(0\right)}\ket{n,n',m}_{s_1,s_2,b}\otimes\ket{n,n'}_{i_1,i_2}
	\nonumber
\end{align} 

\noindent where the $s_j,i_j$-modes denote the signal and idler modes of the $j^{\text{th}}$ down-converter and the $b$-mode represents the SH-mode. The state coefficients in Eq.~\ref{eqn:2.1} above are given by 

\begin{equation}
	C_{n,n',m}^{\left(0\right)} = \left(1 - |z|^{2}\right)\;z^{n+n'}\lambda_{m},
	\label{eqn:2.2}
\end{equation}
where we have used the TMSVS coefficients of Eq.~\ref{eqn:4} setting $z_1=z_2\equiv z$, without loss of generality.  In writing Eq.~\ref{eqn:2.1} we allowed for the possibility of an arbitrary pure state seeding the SH-mode; for the case we are considering (i.e. the SH-mode initially in a vacuum state), the coefficients in Eq.~\ref{eqn:2.2} are given by $\lambda_{m}=\delta_{m,0}$.  The Hamiltonian that drives the three-field interaction is given by

\begin{equation}\label{eqn:trilinear_2}
	\hat{H}_{I} = i\hbar\kappa\big(\hat{a}_{s_1}\hat{a}_{s_2}\hat{b}^{\dagger} - \hat{a}_{s_1}^{\dagger}\hat{a}_{s_2}^{\dagger}\hat{b}\big) = i\hbar\kappa\hat{\mathcal{H}},
\end{equation}

\noindent where $\tau=\kappa t$ is the scaled dimensionless time and where $\hat{b}\;\big(\hat{b}^{\dagger}\big)$ is the annihilation (creation) operator acting on the SH-mode. For a more comprehensive discussion on what physical factors determine $\tau$, see \hyperref[sec:Appendix_C]{Appendix C}.  The full state after SHG is then $\ket{\Psi\left(\tau\right)} = \hat{U}_{T}\ket{\Psi\left(0\right)}$, where the time evolution operator is the usual $\hat{U}_{T} = e^{-\frac{it}{\hbar}\hat{H}_{T}} = e^{\tau\hat{\mathcal{H}}}$.  While an exact solution for the time-evolved state can be worked out numerically, some insight into the short-time state evolution can be gleaned from perturbation theory. The state to $l^{\text{th}}$ order $\mathcal{O}\left(\tau^{l}\right)$, can be written as 

\begin{equation}
	\ket{\Psi_{l}\left(\tau\right) } \simeq \sum_{k=0}^{l}\frac{\tau^{k}}{k!}\hat{\mathcal{H}}^{k}\ket{\Psi\left(0\right)}= \sum_{k=0}^{l}\frac{\tau^{k}}{k!}\ket{\psi_{k}},
	\label{eqn:2.3}
\end{equation}   

\noindent where $\ket{\psi_{0}}=\ket{\Psi\left(0\right)}$ and, by definition, $\ket{\psi_{k+1}} =\hat{\mathcal{H}} \ket{\psi_{k}}$. This can be written in a compact form in terms of the $k^{\text{th}}$-order corrections to the joint-idler subsystem as

\begin{equation}
	\ket{\Psi_{l}\left(\tau\right)} \simeq \sum_{n,n',m}^{\infty}\;\sum_{k=0}^{l}\tau^{k}\ket{\Phi_{n,n',m}^{\left(k\right)}}_{i}\otimes\ket{n,n'}_{s}\otimes\ket{m}_{b}
	\label{eqn:2.4}
\end{equation}

\noindent where the designations $i\to i_1,i_2$ and $s\to s_1,s_2$ have been made for notational convenience; the corrections to the time-evolved state are explicitly worked out in \hyperref[sec:Appendix_A]{Appendix A}. Tracing out the signal modes, we find for the reduced $i_1i_2;b$ system to $l^{\text{th}}$-order

\begin{align}
	&\rho_{i_1,i_2,b}^{\left(l\right)}\left(\tau\right) \simeq \frac{\text{Tr}_{s}\left[\ket{\Psi_{l}\left(\tau\right)}\bra{\Psi_{l}\left(\tau\right)}\right]}{\text{Tr}\left[\ket{\Psi_{l}\left(\tau\right)}\bra{\Psi_{l}\left(\tau\right)}\right]} \label{eqn:2.5} \\
	&\simeq \!\sum_{n,n'}^{\infty}\!\sum_{m,m'}^{\infty}\!\sum_{k+k'=0}^{l}\!\!\!\tau^{k+k'}\ket{\Phi_{n,n',m}^{\left(k\right)}}_{i}\bra{\Phi_{n,n',m'}^{\left(k'\right)}} \otimes\ket{m}_{b}\bra{m'},
	\nonumber
\end{align}

\noindent where $\delta_{q,q'}$ is the usual Kronecker delta function and where the multiplicative factor normalizing the reduced density matrix is understood to be time-dependent. The short-time state statistics can be investigated numerically through computation of the Mandel $Q$ factor given for the $j^{\text{th}}$-mode by

\begin{equation}
	Q_j = \frac{\Delta^{2}\hat{n} - \braket{\hat{n}}}{\braket{\hat{n}}} = \frac{\Delta^{2}\hat{n}}{\braket{\hat{n}}} - 1,
	\label{eqn:Q_fac}
\end{equation}

\begin{figure}[H]
	\centering
	\includegraphics[width=0.95\columnwidth,keepaspectratio]{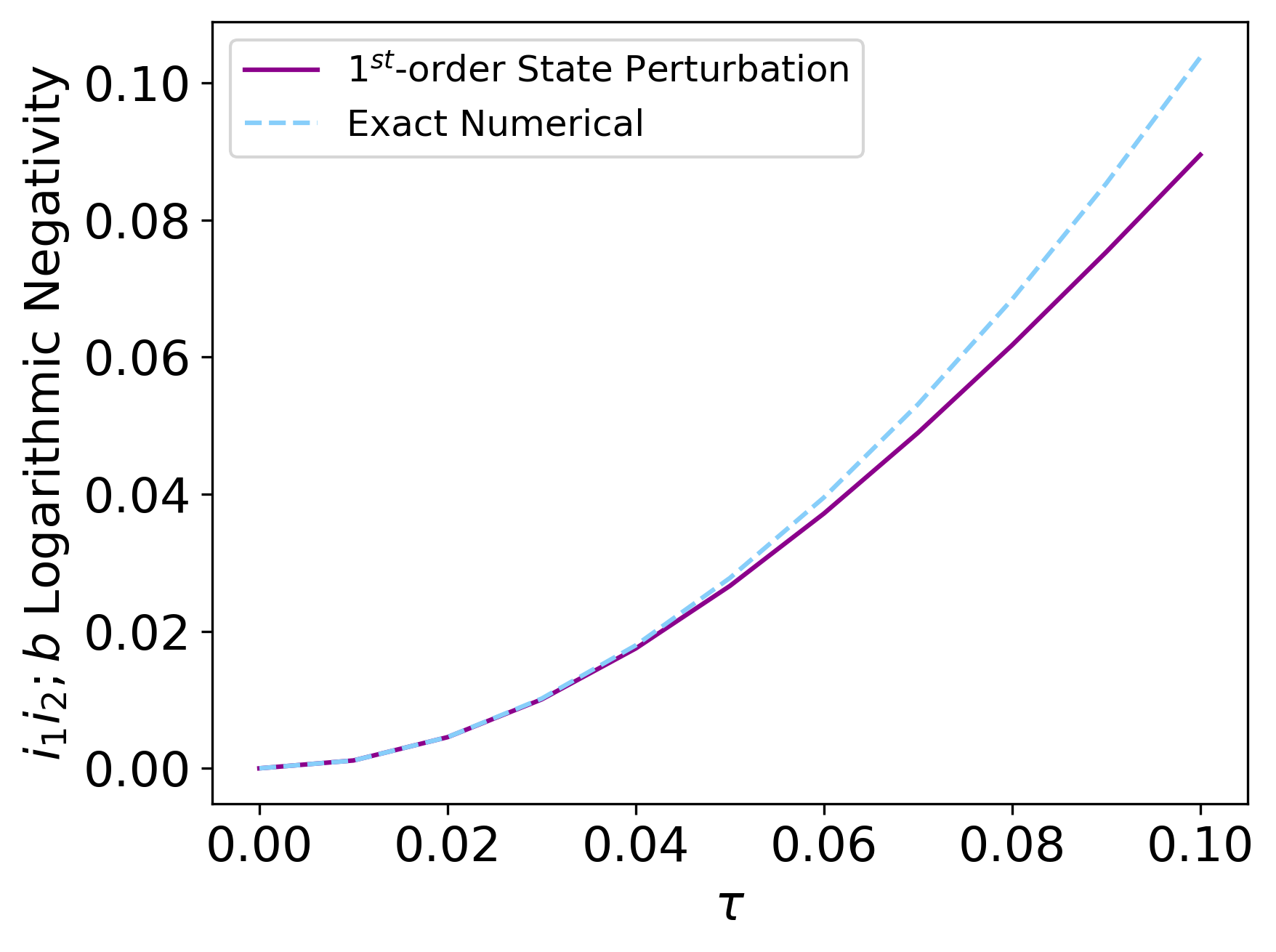}
	\caption{A comparison of the $i_1i_2;b$ logarithmic negativity obtained through full numerical computation of the state and $1^{\text{st}}$-order state perturbation with initial average photon numbers $\bar{n}_{s_1}\left(0\right) = \bar{n}_{s_2}\left(0\right) = 1.5$.}
	\label{fig:2.2}
\end{figure}

\noindent where $Q_j=0$ denotes Poissonian statistics and $Q_j<0,\;Q_j>0$ denotes sub- and super-Poissonian statistics, respectively. At short interaction times, the term in Eq.~\ref{eqn:trilinear_2} responsible for generating a photon at second-harmonic frequency dominates the evolution and the SH-mode begins to thermalize, resulting in a photon number distribution with super-Poissonian photon statistics. Consequently, the signal modes, which remain peaked at the vacuum, becomes \textit{less} super-Poissonian. For very large interaction times, outside what is experimentally accessible with current technologies, the $Q$ factors for both the signal modes as well as the SH-mode fluctuate around a fairly linear increase; thus the modes will ultimately become broader (more super-Poissonian). For short interaction times however, the SH-mode sharply becomes super-Poissonian as the distribution begins to look thermal-like. It is also worth noting that all modes remain Gaussian as a result of this evolution (i.e. all modes have positive Wigner functions). 

\subsection{\label{sec:Entanglement} c.) Entanglement properties of the system}

\begin{figure}[H]
	\centering
	\includegraphics[width=1\linewidth,keepaspectratio]{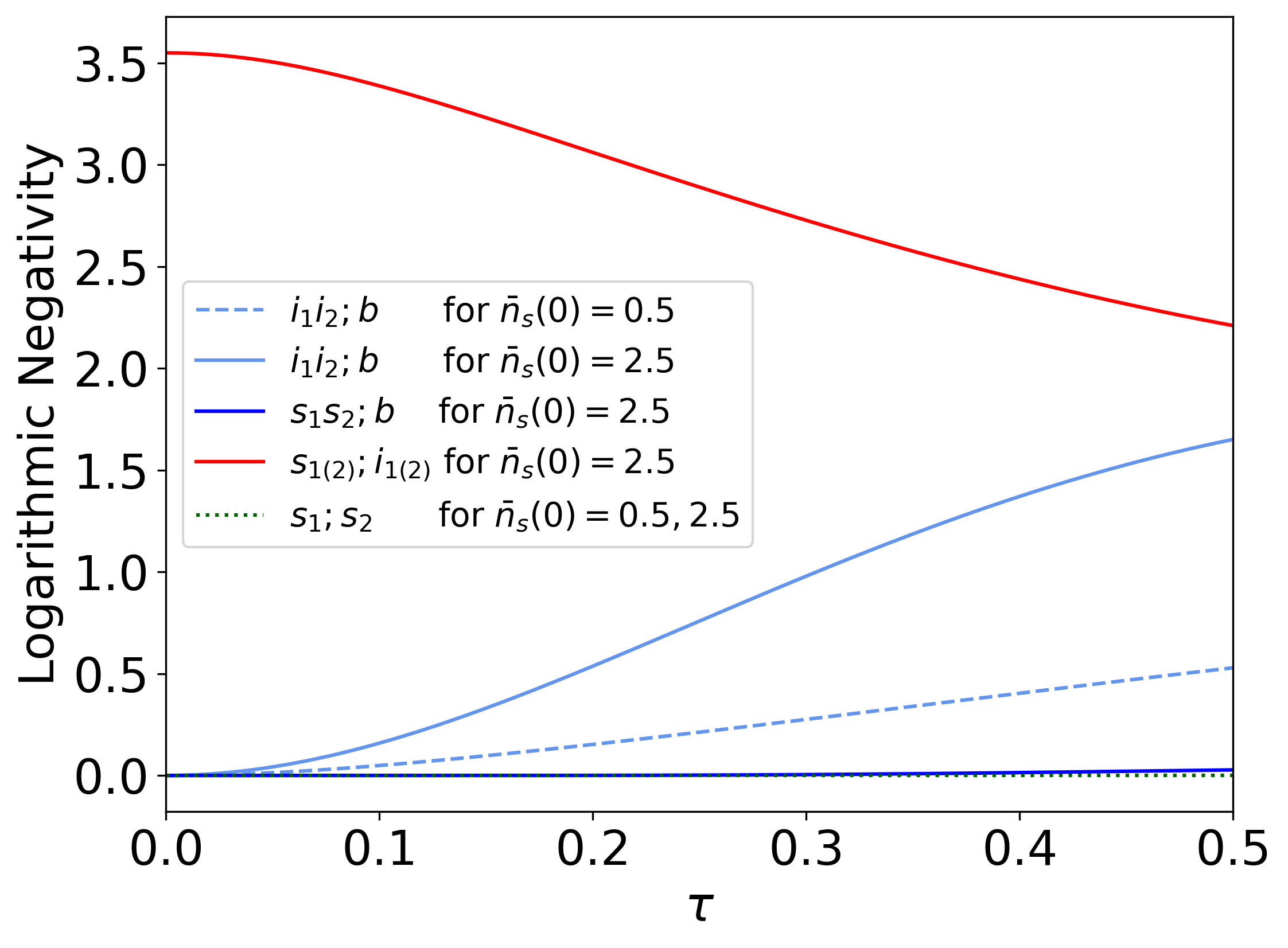}
	\caption{Full numerical computation of the logarithmic negativity for several bipartite divisions of modes for the cases of $\bar{n}_{s_1}\left(0\right) = \bar{n}_{s_2}\left(0\right)=2.5$. The $i_1i_2;b$ entanglement is enhanced with increasing initial average photon numbers as evidenced by the solid-cyan and dashed-cyan curves. Note that for $\bar{n}_s(0)=0.5$, the $s_{1(2)};i_{1(2)}$ entanglement (not shown) remains nearly constant at the value $\approx\text{log}_2 e^{2r_{1(2)}}$ for the time frame of the figure, where $r_{1(2)}$ is the squeeze parameter of the first(second) TMSVS.}
	\label{fig:2.3}
\end{figure}

\noindent As a means of quantifying entanglement between the SH-mode and the joint-idler subsystem, we consider the  logarithmic negativity \cite{ref:AgarwalTextbook}, which for a bipartite division of subsystems $A$ and $B$ is given by 

\begin{equation}
	E_{\mathcal{N}}\left(\rho_{A,B}\right) =\text{log}_{2}\left[1+2\mathcal{N}\left(\rho_{A,B}\right)\right]= \text{log}_{2}||\rho_{A,B}^{T_{B}}||,
	\label{eqn:log1}
\end{equation} 
where $\mathcal{N}\left(\rho_{a,b}\right)$ is the negativity, which stems from the PPT (positive partial transpose) criterion for separability \cite{ref:Simon}, expressed as  

\begin{equation}
	\mathcal{N}(\rho_{A,B}) = \frac{||\rho_{A,B}^{T_{B}}||-1}{2} = \sum_{i}|\lambda_{i}^{(-)}|.
	\label{eqn:log2}
\end{equation}
where $||\rho_{A,B}^{T_{B}}||$ is the trace norm of the partial transpose with respect to the $B$-subsystem of the density operator $\rho_{a,b}$ and where $\lambda_{i}^{\left(-\right)}$ represents the negative eigenvalues of $\rho_{A,B}^{T_{B}}$.   

The negativity, and subsequently the logarithmic negativity, does not increase under local operations and classical communications (LOCC), making it an entanglement monontone \cite{ref:Vidal}. However, the negativity does not constitute an entanglement measure as it is zero for entangled states that have a positive partial transpose. Note though, that entanglement measures are in general $NP$-hard to compute.  For reference, between the signal and idler modes of a TMSVS, there is a logarithmic negativity of $E_{\mathcal{N}}\left(\rho\right) = \text{log}_{2}e^{2r}$ where $r$ is the squeeze parameter, related to the average photon number of the two-mode state by $\bar{n}_{\text{tmsvs}}=2\sinh^{2} r$.  Note that for the pure two-mode squeezed vacuum state, the entanglement can be derived directly from the marginal eigenvalues, where the photon number basis is the Schmidt basis, and the probabilities $P\left(n,n\right)$ from Eq.~\ref{eqn:5} are the Schmidt eigenvalues.  Since the distribution of Schmidt eigenvalues is geometric, the entropy of entanglement can be found to be $E =\cosh^{2}r\times h_2\left(\tanh^2 r\right)$, where $h_2\left(x\right)$ is the binary entropy function \footnote{The binary entropy function $h_{2}(x)$ is given by $-x\log_{2}(x)-(1-x)\log_{2}(1-x)$ and is defined for $x\in[0,1]$.}. In the limit of large squeezing, this becomes $E \approx \log_{2}(e/4) + \log_{2}(e^{2r})$, which is the logarithmic negativity up to a constant offset.

\begin{figure*}
	\includegraphics[width=0.9\linewidth,keepaspectratio]{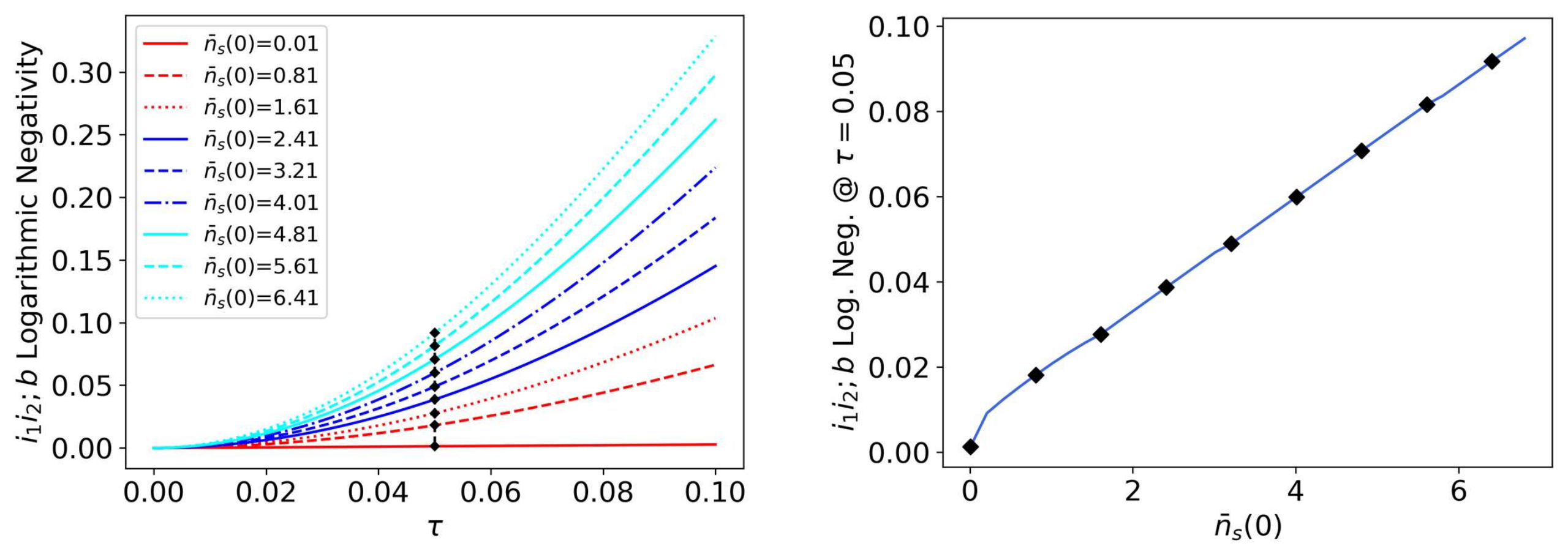}
	\caption{(left) Logarithmic negativity plotted against scaled time $\tau$ for initial signal average photon numbers $\bar{n}_{s_1}(0)=\bar{n}_{s_2}(0)=\bar{n}_{s}(0)=\left\{0,..,7 \right\}$. (right) The logarithmic negativity plotted at time $\tau=0.05$ corresponding to the black dotted points in the left plot. The degree of entanglement as measured by the logarithmic negativity increases nearly linearly (i.e. the negativity increases exponentially) as a function of initial average photon number and persists for feasibly large initial squeezing.}
	\label{fig:LogNeg_Trend}
\end{figure*}

The partial transpose with respect to the SH-mode of the reduced density matrix of Eq.~\ref{eqn:2.5} amounts to flipping the indices on the $b$-mode $\ket{m}_{b}\bra{m'}\to\ket{m'}_{b}\bra{m}$. The partially-transposed reduced density matrix is given explicitly in Eq.~\ref{eqn:a.8a}.  Note that tracing over the SH-mode yields two separable thermal states in the idler modes; to first-order, this corresponds to the simplification $\sum_{m}\alpha_{n,n',m,m}\equiv 0$ and likewise to second order $\sum_{m}\beta_{n,n',m,m}\equiv0$ and $\sum_{m}\Omega_{n,n',m,m}\equiv0$ (see \hyperref[sec:Appendix_A]{Appendix A}).  In fact, this will be true for all corrections to the joint-idler subsystem: the SHG interaction alone is not sufficient enough to entangle the idler modes. We do note that the idler modes can be made to be entangled with a single-photon number resolved (1-PNR) detection, without the need for any local operations, but only for very short interaction times and low initial average photon numbers and with a very low probability of detection. For all intents and purposes, a projective measurement alone will not result in entanglement between idler modes. We briefly discuss in \hyperref[sec:Appendix_B]{Appendix B} a means of remotely entangling the idler modes, through a projective measurement on the SH-mode, however this requires beamsplitting the signal modes, effectively introducing correlations between the two down-converters prior to performing SHG.   \\

The $i_1i_2;b$ logarithmic negativity is plotted in Fig.~\ref{fig:2.2} using first-order perturbation for initial average photon numbers $\bar{n}_{s_1}\left(0\right) = \bar{n}_{s_2}\left(0\right)\equiv\bar{n}_s = 0.5$. Interestingly, first-order perturbation is sufficient to accurately witness-, and capture the trend of-, the $i_1i_2;b$ entanglement for fairly reasonable interaction times of around $\tau \sim 0.04$. We plot the logarithmic negativity based on a full numerical computation of the state in Fig.~\ref{fig:2.3} for different values of initial average photon numbers and for several different bipartite divisions of modes. We find the peak of the $i_1i_2;b$ logarithmic negativity very nearly corresponds to the minimum in the $s_{1\left(2\right)};i_{1\left(2\right)}$ logarithmic negativity (not explicitly shown in Fig.~\ref{fig:2.3}); this can be interpreted as an example of entanglement redistribution as the inherent entanglement between the two modes of a TMSVS is leveraged to entangle the generated second-harmonic field with the non-interacting spatially-separated joint-idler modes. Note from Fig.~\ref{fig:2.3} that the entanglement is not perfectly redistributed, however a measure of entanglement (rather than a monotone) may yet show full equivalence.  The drawback to this approach is that measures are $NP$-hard to compute in general.  An argument can be made on the interplay of entanglement between these two bipartitions by defining the entanglement measure between these two subsystems as $E_{\tau}\left(i_1i_2|s_1s_2b\right)$. Since the trilinear Hamiltonian is local with respect to the $s_1,s_2,b$-modes, the entanglement between this bipartition must remain constant, i.e.

\begin{equation}
	E_{\tau}\left(i_1i_2|s_1s_2b\right) = \text{const.} \;\;\;\;\;\; \forall\;\tau.
	\label{eqn:JS_added_1}
\end{equation}

\noindent Given the initial state is a product of two TMSVSs, we can expand Eq.~\ref{eqn:JS_added_1} at $\tau=0$, where $E_{\tau=0}\equiv E$, accordingly

\begin{align}
	E\left(i_1i_2|s_1s_2b\right) &= E\left(i_1i_2|b\right) + E\left(i_1i_2|s_1s_2\right)  \nonumber \\
	 &= E\left(i_1i_2|b\right) +E\left(i_1|s_1\right) +E\left(i_2|s_2\right) \nonumber \\
	 &= E\left(i_1i_2|b\right) +2E\left(i_1|s_1\right),
	\label{eqn:JS_added_2}
\end{align}

\noindent where here, these approximate relations come from the hypothetical entanglement measure $E_{\tau}$ being additive over tensor products (the state factors this way at $\tau=0$), and many entanglement measures have this property. To the extent that this relation holds for  later times $\tau>0$ will depend on how well the correlations can be decomposed in this way. From this, a reasonable yet approximate argument can be made as to why the logarithmic negativity for the two bipartitions in the last line of Eq.~\ref{eqn:JS_added_2} appear strongly correlated in Fig.~\ref{fig:2.3}.  Most notably is that the $i_1i_2;b$ entanglement is also enhanced by increasing values of initial average photon numbers.  In Fig.~\ref{fig:LogNeg_Trend} we plot the $i_1i_2;b$ logarithmic negativity as a function of scaled time $\tau$ for different values of initial signal average photon number up to $\bar{n}_{s_{1(2)}}=7$ and show how the logarithmic negativity increases near-linearly as a function of initial average photon number.  Stated differently, the negativity grows exponentially as a function of initial average photon number.  For low average photon numbers these two bipartitions constitute the only entanglement within the system, while for larger initial averages, entanglement between the joint-signal modes and the SH-mode begins to form at shorter times, however this entanglement remains small relative to the $i_1i_2;b$ and $s_{1(2)};i_{1(2)}$.

The counter-intuitive nature of this $i_{1}i_{2};b$  entanglement formation should not be overlooked. Interaction between the signals and SH-mode, initially in a vacuum state, does \textit{not} produce entanglement between the signal modes, between the joint-signal and SH-modes (for short times), nor between either signal mode and the SH-mode. Instead, the entanglement that forms is between the SH-mode and the non-interacting idler modes. Interestingly, for no times $\tau$ nor for any value of initial average photon number will the SH-mode become entangled with either idler mode individually, i.e., entanglement $i_{1(2)};b$ never forms, at least as indicated by the negativity which cannot identify entanglement for states with a positive partial transpose. 

\begin{figure}
	\centering
	\includegraphics[width=1\linewidth,keepaspectratio]{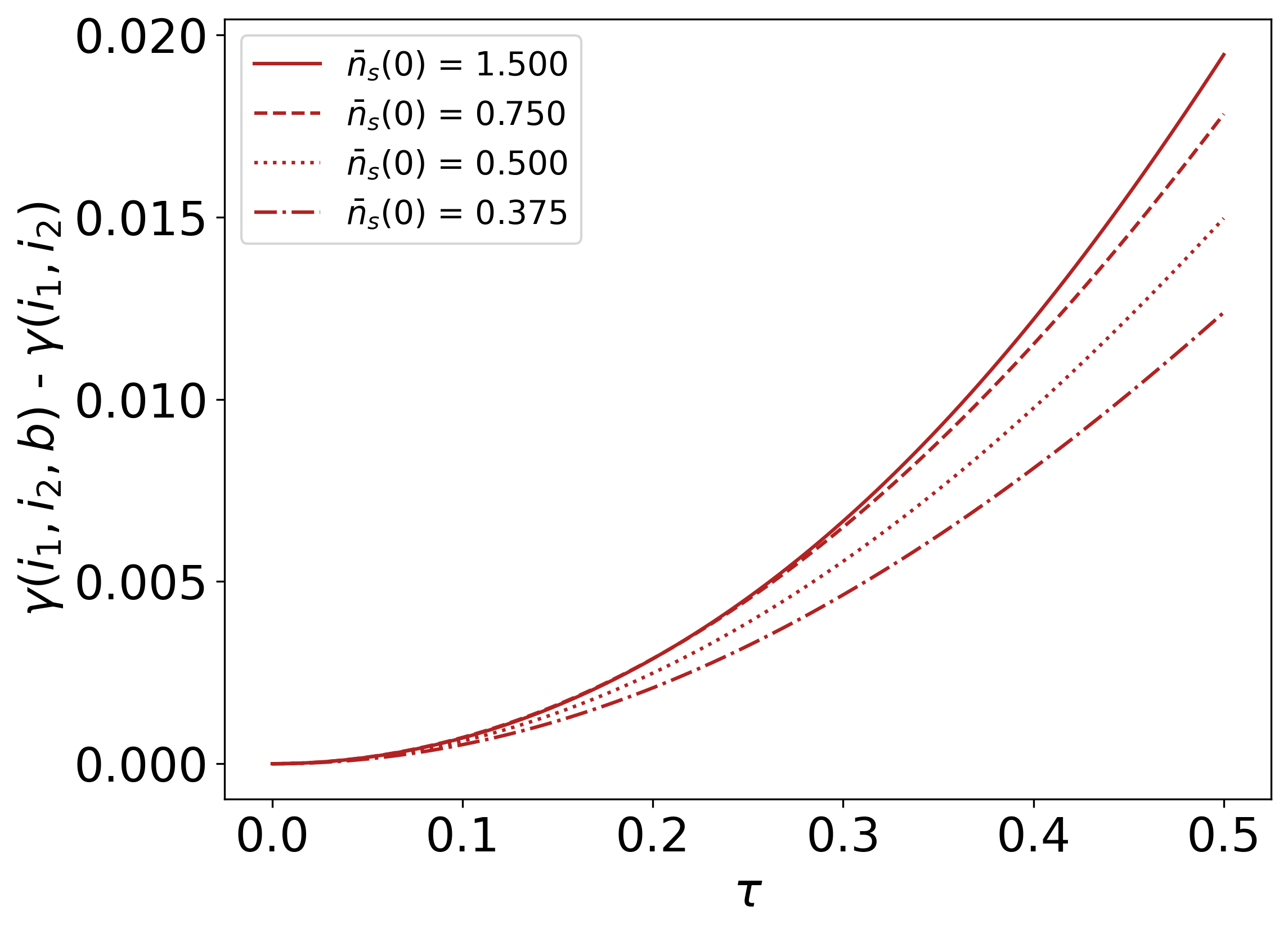}
	\caption{Purity Difference between reduced density matricies $\rho_{i_1,i_2,b}$ and $\rho_{i_1,i_2}$ for several different initial average photon numbers $\bar{n}_{s}\left(0\right)$.  Values $>0$ indicate entanglement between the joint-idler system and the SH-mode. For short time evolution, larger initial averages correspond to greater purity differences.}
	\label{fig:2.4}
\end{figure}

Other entanglement metrics have successfully been able to detect the $i_1i_2;b$ entanglement such as the reduction criterion \cite{ref:Horodecki} which requires computation of the spectrum for the quantity $\rho_{i_1,i_2}\left(\tau\right) - \rho_{i_1,i_2,b}\left(\tau\right)$; $i_1i_2;b$ entanglement is necessitated by the presence of negative eigenvalues, which occur for all times $\tau>0$.  Similarly we consider the difference in purity between the two reduced systems $\gamma\left(i_1,i_2,b\right)-\gamma\left(i_1,i_2\right)$, where the purity is given by the usual $\gamma\left(j\right)=\text{Tr}\left[\rho_{j}^{2}\right]$, because determining the purity requires fewer resources than the reduction or PPT criterion. Note that by virtue of the initial state being composed of a tensor product of two TMSVS, the reduced density matrix $\rho_{i_1,i_2,b}$ already represents a mixed state; however, the (positive) non-zero difference seen for increasing values of $\bar{n}_{s_1}\left(0\right)=\bar{n}_{s_2}\left(0\right)=\bar{n}_{s}\left(0\right)$ in Fig.~\ref{fig:2.4} indicates a higher degree of mixedness for the joint-idler system when tracing out the SH-mode.  This is sufficient enough to conclude entanglement between the SH-mode and the joint-idler system. Note that when it is straightforward to obtain and compute the eigenvalues of the partial transpose, the negativity will encompass all states whose entanglement is witnessed by the purity difference \cite{schneeloch2020negativity}. However this is at the cost of requiring more resources to determine.

\section{\label{sec:Conclusion} III. Conclusion}

\noindent We have demonstrated a form of passive remote entanglement using continuous-variable (CV) states of light utilizing second-order nonlinear interactions whose evolution is described by the trilinear Hamiltonian, that does not require the implementation of any state-reductive measurements. More specifically, our scheme allows for the formation of entanglement between the generated second-harmonic field and the non-interacting spatially-separated joint-idler fields.  We have verified this entanglement through both a perturbative treatment of the state evolution as well as through full numerical calculation of the state. We note this entanglement persists for experimentally accessible interaction times and scales reasonably with initial average photon numbers seeding the interaction.\\

\section{IV. Acknowledgements}\label{sec:Acknowledgements}

\noindent RJB acknowledges support from the NRC RAP.  CCG acknowledges support under AFRL Visiting Faculty Fellowship Program, AFRL Contract No. SA-10032021060386.  JS, CCT, MLF and PMA acknowledge support from the Air Force Office of Scientific Research. 

The views expressed are those of the authors and do not reflect the official guidance or position of the United States Government, the Department of Defense or of the United States Air Force. The appearance of external hyperlinks does not constitute endorsement by the United States Department of Defense (DoD) of the linked websites, or of the information, products, or services contained therein. The DoD does not exercise any editorial, security, or other control over the information you may find at these locations. \\

\section{Appendix A: Short-time trilinear state evolution via perturbation}\label{sec:Appendix_A}

\noindent Here we will provide a more thorough description of the perturbative approximation employed in describing the state evolution driven by the trilinear Hamiltonian of Eq.~\ref{eqn:trilinear_2}. The time-dependent state is given by $\ket{\Psi\left(\tau\right)} = \hat{U}_{T}\ket{\Psi\left(0\right)}$, where $\tau=\kappa t$ and where the time evolution operator can be expanded out as 

\begin{equation}
	\hat{U}_{T} = e^{-\frac{it}{\hbar}\hat{H}_{T}} = e^{\tau\hat{\mathcal{H}}}\simeq 1+\tau\mathcal{H} + \frac{1}{2!}\tau^{2}\mathcal{H}^{2}+..\;.
	\label{eqn:a.1}
\end{equation} 

Action of the Hamiltonian $\mathcal{H}$ on an arbitrary term comprising the initial state Eq.~\ref{eqn:2.1} can be worked out straight-forwardly as

\begin{widetext}
	\begin{align}
		C_{n,n',m}^{\left(0\right)}\mathcal{H}\ket{n,n',m}_{s,b}\otimes\ket{n,n'}_{i} &= \sqrt{nn'\left(m+1\right)}\;	C_{n,n',m}^{\left(0\right)}\ket{n-1,n'-1,m+1}_{s,b}\ket{n,n'}_{i} -  \nonumber \\
		& \;\;\;\;\;\;\;\;\;\;\;\;\;\;\;\;\;\;\;\;\;\;\;\;\;\;\;\;- \sqrt{\left(n+1\right)\left(n'+1\right)m}\;	C_{n,n',m}^{\left(0\right)}\ket{n+1,n'+1,m-1}_{s,b}\ket{n,n'}_{i} \nonumber \\
		&= \sqrt{\left(n+1\right)\left(n'+1\right)m}\;	C_{n+1,n'+1,m-1}^{\left(0\right)}\ket{n,n',m}_{s,b}\ket{n+1,n'+1}_{i} -  \nonumber \\
		& \;\;\;\;\;\;\;\;\;\;\;\;\;\;\;\;\;\;\;\;\;\;\;\;\;\;\;\;- \sqrt{nn'\left(m+1\right)}\;	C_{n-1,n'-1,m+1}^{\left(0\right)}\ket{n,n',m}_{s,b}\ket{n-1,n'-1}_{i} \nonumber \\
		&= \Big[\sqrt{\left(n+1\right)\left(n'+1\right)m}\;	C_{n+1,n'+1,m-1}^{\left(0\right)}\ket{n+1,n'+1}_{i} - \Big. \nonumber \\
		& \Big.\;\;\;\;\;\;\;\;\;\;\;\;\;\;\;\;\;\;\;\;\;\;\;\;\;\;\;\;- \sqrt{nn'\left(m+1\right)}\; C_{n-1,n'-1,m+1}^{\left(0\right)}\ket{n-1,n'-1}_{i}\Big]\otimes\ket{n,n',m}_{s,b} \nonumber \\
		&= \ket{\Phi_{n,n',m}^{\left(1\right)}}_{i}\otimes\ket{n,n',m}_{s,b}, 
		\label{eqn:a.2}
	\end{align} 
\end{widetext}
where the designations $i\to i_1,i_2$ and $s\to s_1,s_2$ have been made for notational convenience and where the term in the square brackets constitutes the first-order correction to the joint-idler modes for the initial state

\begin{widetext}
	\begin{align}
		\ket{\Phi_{n,n',m}^{\left(1\right)}}_{i} &=  \sqrt{\left(n+1\right)\left(n'+1\right)m}\;	C_{n+1,n'+1,m-1}^{\left(0\right)}\ket{n+1,n'+1}_{i} - \sqrt{nn'\left(m+1\right)}\; C_{n-1,n'-1,m+1}^{\left(0\right)}\ket{n-1,n'-1}_{i}, \nonumber \\
		&= A_{n,n',m}\ket{n+1,n'+1}_{i} +  B_{n,n',m}\ket{n-1,n'-1}_{i}.
		\label{eqn:a.4}
	\end{align}
\end{widetext}

In working out the state perturbatively, we note that all higher-order corrections can be cast in terms of the first-order corrections derived in Eq.~\ref{eqn:a.4}.  Following Eq.~\ref{eqn:2.3}, and utilizing the relation $\ket{\psi_{k+1}} = \mathcal{H}\ket{\psi_{k}}$ we can write

\begin{align}
	C_{n,n',m}^{\left(0\right)}\mathcal{H}^{2}\ket{n,n',m}_{s,b}\ket{n,n'}_{i} &=\mathcal{H}\ket{\Phi_{n,n',m}^{\left(1\right)}}_{i}  \ket{n,n',m}_{s,b} \nonumber \\
	&=\ket{\Phi_{n,n',m}^{\left(2\right)}}_{i}  \ket{n,n',m}_{s,b},
	\label{eqn:a.5}
\end{align}  
where the second-order correction can now be written in terms of the first-order correction as

\begin{align}
	\ket{\Phi_{n,n',m}^{\left(2\right)}}_{i} = &\sqrt{\left(n+1\right)\left(n'+1\right)m}\ket{\Phi_{n+1,n'+1,m-1}^{\left(1\right)}}_{i} - \nonumber \\
	& -  \sqrt{nn'\left(m+1\right)}\ket{\Phi_{n-1,n'-1,m+1}^{\left(1\right)}}_{i}.
	\label{eqn:a.6}
\end{align}

In similar fashion the third order correction can be found via $\mathcal{H}\ket{\Phi_{n,n',m}^{\left(2\right)}}_{i}\ket{n,n',m}_{s,b} = \ket{\Phi_{n,n',m}^{\left(3\right)}}_{i}\ket{n,n',m}_{s,b}$, or more generally

\begin{equation}
	\mathcal{H}\ket{\Phi_{n,n',m}^{\left(k\right)}}_{i}\ket{n,n',m}_{s,b} = \ket{\Phi_{n,n',m}^{\left(k+1\right)}}_{i}\ket{n,n',m}_{s,b}.
	\label{eqn:a.7}
\end{equation}

Combining terms, we write the time-evolved state to $l^{\text{th}}$-order in the compact form

\begin{equation}
	\ket{\Psi_{l}\left(\tau\right)} \simeq \sum_{n,n',m}^{\infty}\;\sum_{k=0}^{l}\tau^{k}\ket{\Phi_{n,n',m}^{\left(k\right)}}_{i}\otimes\ket{n,n',m}_{s,b},
	\label{eqn:a.8}
\end{equation}
where the zeroth-order correction is clearly given by $\ket{\Phi_{n,n',m}^{\left(0\right)}}_{i}=C_{n,n',m}^{\left(0\right)}\ket{n,n'}_i$. This is leading to the derivation of the perturbed state of Eq.~\ref{eqn:2.4}.  For first-order corrections to the state, we can work out the partial-transposed reduced density matrix in the Fock basis to find

\begin{widetext}
	\begin{align}
			\rho_{i_1,i_2,b}^{\left(1\right)\;T_{b}}\left(\tau\right) &\simeq \sum_{n,n'}^{\infty}\sum_{m,m'}^{\infty}\left[\ket{\Phi_{n,n',m}^{\left(0\right)}}_{i}\bra{\Phi_{n,n',m'}^{\left(0\right)}} +  \tau\left(\ket{\Phi_{n,n',m}^{\left(1\right)}}_{i}\bra{\Phi_{n,n',m'}^{\left(0\right)}} + \ket{\Phi_{n,n',m}^{\left(0\right)}}_{i}\bra{\Phi_{n,n',m'}^{\left(1\right)}}\right) +  \right. \nonumber \\
			&\left. \;\;\;\;\;\;\;\;\;\;\;\;\;\;\;\;\;\;\;\;\;\;\;\;\;\;\;\;\;\;\;\;\;\;\;\;\;\;\;\;\;\;\;\;\;\;\;\;\;\;\;\;\;\;\;\;\;\;\;\;\;\;\;\;\;\;\;\;\;\;\;\;\;\;\;\;\;\;\;\;\;\;\;\;\;\;\;\;\;\;\;\;+  \tau^{2} \ket{\Phi_{n,n',m}^{\left(1\right)}}_{i}\bra{\Phi_{n,n',m'}^{\left(1\right)}} \right]\otimes\ket{m'}_{b}\bra{m} \nonumber \\
			&\simeq \sum_{n,n'}^{\infty}\sum_{m,m'}^{\infty}\Big[\Big(C_{n,n',m'}^{\left(0\right)\; *}C_{n,n',m}^{\left(0\right)} + \tau^{2}\Omega_{n,n',m,m'}\Big)\ket{n,n'}_{i}\bra{n,n'} + \tau\Big(\alpha_{n,n',m,m'}\ket{n+1,n'+1}_{i}\bra{n,n'} + \Big.\Big. \nonumber\\
			 &\Big.\Big.\;\;\;\;\;\;\;\;\;\;\;\;\;\;\;\;\;\;\;\;\;\;\;\;\;\;\;\;\;\;\;\;\;\;\;\;+\alpha_{n,n',m',m}^{*}\ket{n,n'}_{i}\bra{n+1,n'+1}\Big) + \tau^2\Big(\beta_{n,n',m,m'}\ket{n+2,n'+2}_{i}\bra{n,n'} + \Big.\Big. \nonumber \\
			 &\Big.\Big.\;\;\;\;\;\;\;\;\;\;\;\;\;\;\;\;\;\;\;\;\;\;\;\;\;\;\;\;\;\;\;\;\;\;\;\;\;\;\;\;\;\;\;\;\;\;\;\;\;\;\;\;\;\;\;\;\;\;\;\;\;\;\;\;\;\;\;\;\;\;\;\;\;\;\;+ \beta^{*}_{n,n',m',m}\ket{n,n'}_{i}\bra{n+2,n'+2}\Big)\Big]\otimes\ket{m'}_{b}\bra{m},
			\label{eqn:a.8a}
	\end{align}

where the correction coefficients are given by

\begin{align}
	\Omega_{n,n',m,m'} &= A_{n-1,n'-1,m}A_{n-1,n'-1,m'}^{*} + +B_{n+1,n'+1,m}^{*}B_{n+1,n'+1,m'}^{*}\nonumber \\
	\alpha_{n,n',m,m'} &= A_{n,n',m}C^{\left(0\right)\;*}_{n,n',m'} + B_{n,n',m'}^{*}C^{\left(0\right)}_{n,n',m} \nonumber\\
	\beta_{n,n',m,m'} &=  A_{n+1,n'+1,m}B^{*}_{n+1,n'+1,m'}.
	\label{eqn:a.8b}
\end{align}
\end{widetext}

Note that second order corrections of the form $\propto\ket{\Phi_{n,n',m'}^{\left(2\right)}}_{i}\bra{\Phi_{n,n',m'}^{\left(0\right)}}$ and its Hermitian conjugate do not contribute to the logarithmic negativity and have thus been disregarded in writing Eq.~\ref{eqn:a.8a}.  This is the form of the state used in the approximation of the logarithmic negativity and purity-difference plots in the main body of the text. \\

\section{Appendix B: Remote Entanglement between idler modes}\label{sec:Appendix_B}

\noindent Usual methods of CV entanglement swapping that employ two SPDC sources require one mode (signal) from each source to first fall upon a beamsplitter. Subsequently, opposite quadratures are measured from each output port of the beamsplitter, $\{\hat{x}_{s_1},\hat{p}_{s_2}\}$, and then one of the remaining (idler) modes is displaced by $\beta=\hat{x}_{s_1}+i\hat{p}_{s_2}$. This results in the teleportation of the initial SPDC entanglement. Such a scheme is typically used in conjunction with continuous-variable measurement-device-independent quantum-key-distribution (cv-mdi-qkd) protocols \cite{ref:Li}. More recently, a similar scheme was proposed using two-mode squeezed coherent states \cite{ref:Kumar}. They went on to find that photon subtraction of the source states dramatically enhanced transmission distances while suffering only a slight decrease in the maximum achievable secure key rate.  

 \begin{figure}[H]
	\centering
	\includegraphics[width=1.0\linewidth,keepaspectratio]{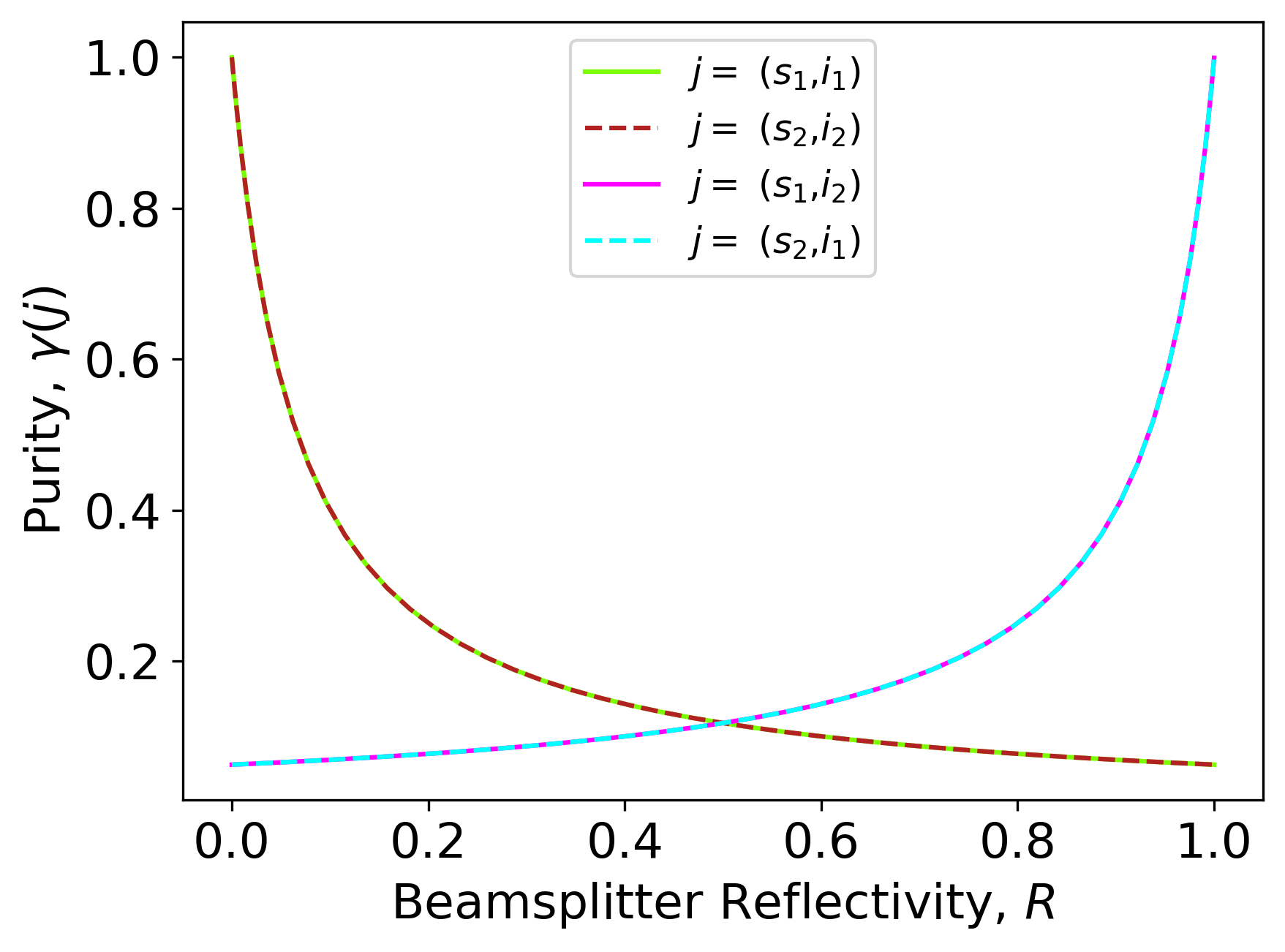}
	\caption{Purity $\gamma(j)$ for several different reduced density matrices of the composite $\ket{\xi_1}_{s_1,i_1}\otimes\ket{\xi_2}_{s_2,i_2}$ system after beam splitting of the signal modes. Note that for a reflectivity $R=\sin^{2}\tfrac{\theta}{4}=0.5$, corresponding to a beam splitter angle of $\theta=\pi/2$, the reduced density matrix $\rho_{s_{1(2)}i_{1(2)}}$ is as mixed as $\rho_{s_1i_2}$ and $\rho_{s_2i_1}$. The average photon number for each signal mode is $\bar{n}_{s}\left(0\right) = 1.5$}
	\label{fig:3.4}
\end{figure}

Other works investigating CV entanglement swapping using SPDC sources are: Jia \textit{et al.} \cite{ref:Jia} considered the use of two nondegenerate optical parametric amplifiers (OPAs) and employed a Bell-state measurement on one mode from each source, entangling the remaining optical modes. They report quantum correlation degrees of 1.3dB (1.12dB) below the standard quantum limit for the amplitude (phase) quadratures resulting from this unconditional entanglement formation. Further, Takei \textit{et al.} \cite{ref:Takei} used EPR states generated via beamsplitting two opposite-quadrature-squeezed optical parametric oscillators (the output of which is the single-mode squeezed vacuum state; upon beam splitting, the two-mode state is the two-mode squeezed vacuum state) to demonstrate teleportation of entanglement. Their detection scheme likewise employed Bell-state measurements. For a more comprehensive reference on CV entanglement swapping, see Marshall and Weedbrook \cite{ref:Marshall}.

\begin{figure*}
	\includegraphics[width=0.8\linewidth,keepaspectratio]{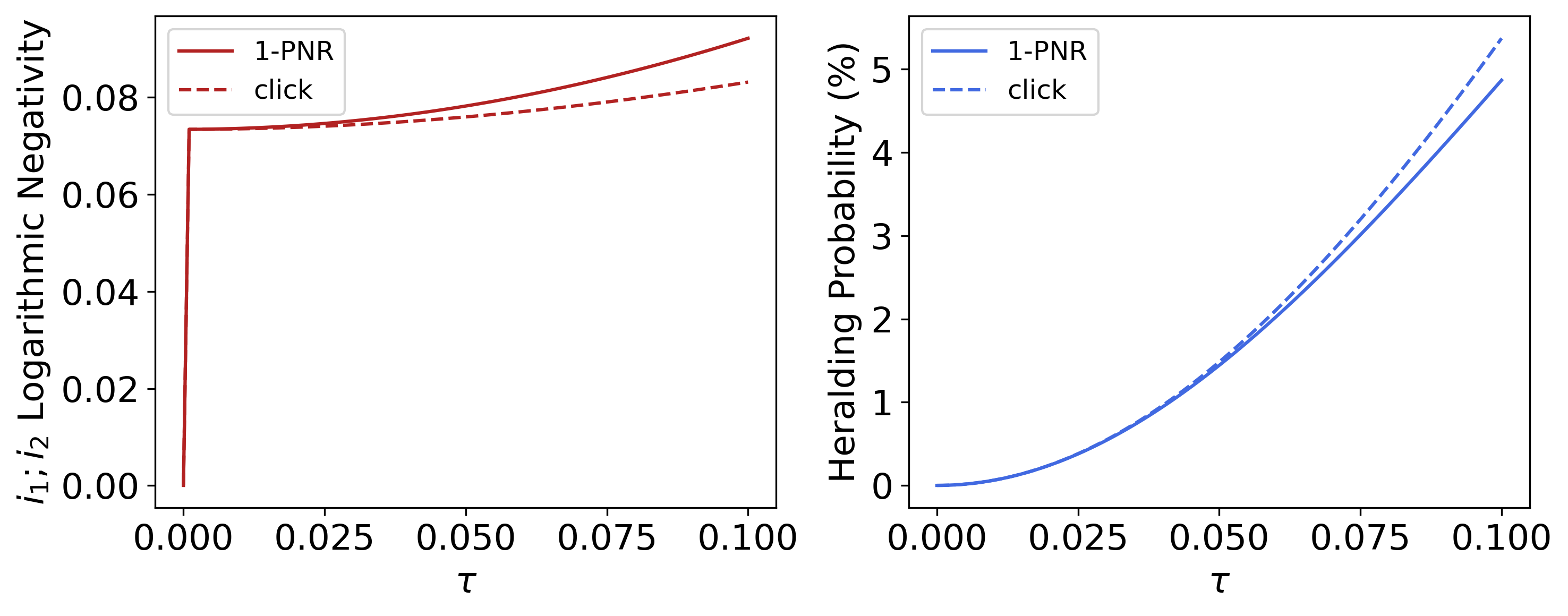}
	\caption{(left) Logarithmic negativity between $i_1;i_2$ modes and (right) the corresponding heralding probability.  Both figures are based on a numerical computation of the state evolution for $\bar{n}_{s_1}\left(0\right) = \bar{n}_{s_2}\left(0\right) = 2.5$ for both click-detection (dashed) and 1-PNR detection (solid).}
	\label{fig:3.3}
\end{figure*}

While our proposed scheme for $i_1,i_2;b$ entanglement does not result in true entanglement swapping since the modes that are subsequently entangled are not teleported as a consequence of this interaction, remote entanglement does form between the joint-idler and SH-modes, as discussed in Section \hyperref[sec:RemoteEntanglement]{II\ref*{sec:RemoteEntanglement}}. A natural extension to the aforementioned entanglement formation would be to apply our system to a well known CV entanglement swapping protocol that uses SPDC sources \cite{ref:Li}\cite{ref:Kumar}.   To start, consider a 50:50 $\hat{J}_{y}$-type beam splitter $\hat{B}_{y}$ \cite{ref:Yurke} (which can be realized experimentally using a Mach-Zehnder interferometer with a variable phase between beam paths) with a scattering matrix resulting in the Boson operator transformation

 \begin{align}
 	\vec{\hat{a}}_{\text{out}} &= \hat{B}_{y}^{\dagger}\vec{\hat{a}}_{\text{in}}\hat{B}_{y} = \hat{U}\;\vec{\hat{a}}_{\text{in}} \;\to\; \nonumber \\
 	&\;\;\;\;\;\;\;\;\;\;\;\;\;\;\to\;
 	\begin{bmatrix}
 		\hat{a}_{s_1}  \\
 		\hat{a}_{s_2} 
 	\end{bmatrix}_{\text{out}} 
 	=
 	\frac{1}{\sqrt{2}}
 	\begin{bmatrix}
 		1 & 1 \\
 		-1 & 1
 	\end{bmatrix} 
 	\begin{bmatrix}
 		\hat{a}_{s_1}  \\
 		\hat{a}_{s_2} 
 	\end{bmatrix}_{\text{in}},
 	\label{eqn:3.1}
 \end{align}
 
\noindent resulting in a transformation of the trilinear Hamiltonian of Eq.~\ref{eqn:trilinear_2} given by 
 
\begin{align}
	\hat{H}_{I}^{'} &= \hat{B}_{y}^{\dagger}\hat{H}_{I}\hat{B}_{y} = i\hbar\kappa\big(\hat{B}_{y}^{\dagger}\hat{a}_{s_1}\hat{a}_{s_2}\hat{B}_{y}\hat{b}^{\dagger} - \hat{B}_{y}^{\dagger}\hat{a}_{s_1}^{\dagger}\hat{a}_{s_2}^{\dagger}\hat{B}_{y}\hat{b}\big) \nonumber \\
	&= i\hbar\kappa\big(\hat{B}_{y}^{\dagger}\hat{a}_{s_1}\hat{B}_{y}\hat{B}_{y}^{\dagger}\hat{a}_{s_2}\hat{B}_{y}\hat{b}^{\dagger} - \hat{B}_{y}^{\dagger}\hat{a}_{s_1}^{\dagger}\hat{B}_{y}\hat{B}_{y}^{\dagger}\hat{a}_{s_2}^{\dagger}\hat{B}_{y}\hat{b}\big) \nonumber \\
	&= i\hbar\frac{\kappa}{2}\left(\left(\hat{a}_{s_2}^{2} - \hat{a}_{s_1}^{2}\right)\hat{b}^{\dagger} + \left(\hat{a}_{s_1}^{\dagger\;2} - \hat{a}_{s_2}^{\dagger\;2}\right)\hat{b}\right) \nonumber \\
	&= \hat{H}_{\text{shg}}^{(s_2)} - \hat{H}_{\text{shg}}^{(s_1)}, \label{eqn:3.2}
\end{align} 
where $\hat{H}_{\text{shg}}^{\left(j\right)} = i\hbar\kappa\left(\hat{a}_{j}^{2}\hat{b}^{\dagger} - \hat{a}_{j}^{\dagger\;2}\hat{b}\right)$ is the degenerate SHG Hamiltonian in which two photons are annihilated from the $j$-mode to create a second-harmonic photon (and conjugate operation corresponding to the reverse process).  This makes Eq.~\ref{eqn:3.2} a sum of degenerate SHG terms each with $\kappa/2$ efficiency.  This transformation is realized physically by beam splitting the signal modes prior to seeding the nonlinear crystal such that 

\begin{align}
	\ket{\Psi\left(t\right)} = e^{-i\tfrac{t}{\hbar}\hat{H}_{I}}\hat{B}_{y}\ket{\Psi\left(0\right)} &= \hat{B}_{y}\hat{B}_{y}^{\dagger}e^{-i\tfrac{t}{\hbar}\hat{H}_{I}}\hat{B}_{y}\ket{\Psi\left(0\right)} \nonumber \\
	&=  \hat{B}_{y}e^{-i\tfrac{t}{\hbar}\hat{B}_{y}^{\dagger}\hat{H}_{I}\hat{B}_{y}}\ket{\Psi\left(0\right)} \nonumber \\
	&=  \hat{B}_{y}e^{-i\tfrac{t}{\hbar}\hat{H}_{I}^{'}}\ket{\Psi\left(0\right)},
	\label{eqn:added_1}
\end{align}
or $\hat{B}_{y}^{\dagger}\ket{\Psi\left(t\right)} = e^{-i\tfrac{t}{\hbar}\hat{H}_{I}^{'}}\ket{\Psi\left(0\right)}$. This amounts to placing a beam splitter between the output signal modes, which in turn does not affect entanglement formation between the idler modes (but \textit{does} affect the entanglement between $s_{1(2)};i_{1(2)}$).  For this reason, we neglect the beam splitter operation in the last line of Eq.~\ref{eqn:added_1} and instead work with the transformed Hamiltonian of Eq.~\ref{eqn:3.2}, in our analysis.  

The effect of initially beam splitting the signal modes prior to SHG is to create cross correlations between the two down-converters.  More specifically, for a 50:50 beam splitter, entanglement is formed between the two TMSVSs in the form of $s_1i_1;s_2i_2$ entanglement and $s_1i_2;s_2i_1$ entanglement.  Note that neither the signal nor idlers will be entangled after beam splitting.  This can be demonstrated via computation of the purity as shown in Fig.~\ref{fig:3.4}. 

For a fully transmitting beam splitter, tracing out either of the TMSVSs results in unit purity, as the TMSVS is pure and initially not entangled with the second TMSVS.  However, once the signals mix at the beam splitter, tracing out one TMSVS results in a mixed reduced density matrix for the remaining TMSVS. While the beam splitter will not create entanglement between the signal modes nor the idler modes, it \textit{does} create entanglement between the composite states of the individual down-converters. This should not be confused with the classical correlations that beam splitting creates via computation of the state covariances. 

For the case of equal initial averages, no correlations exist between the signal modes nor between the idler modes for any beam splitter angle. However signal mode correlations ($s_1;s_2$) will manifest for unequal initial averages such that $|\text{cov}\big(\hat{X}_{s_1},\hat{X}_{s_2}\big)|= |\text{cov}\big(\hat{Y}_{s_1},\hat{Y}_{s_2}\big)| = |\bar{n}_{s_1}-\bar{n}_{s_2}|$ for a 50:50 $\hat{J}_y$-type beam splitter. Here, $\hat{X}_k,\;\hat{Y}_k$ are the usual quadrature operators $\hat{X}_k = \tfrac{1}{2}\big(\hat{a}_k +\hat{a}^{\dagger}_k\big)$ and $\hat{Y}_k = \tfrac{i}{2}\big(\hat{a}_k -\hat{a}^{\dagger}_k\big)$, respectively. \\

We are interested in investigating the potential entanglement formation between the spatially-separated non-interacting idler modes upon performing a state-projective measurement on the SH-mode.  We consider both a `click' detection corresponding to the projection operator $\hat{P}_{C}=\sum_{n=1}^{\infty}\ket{n}_{b}\bra{n}$ as well as a 1-PNR detection with corresponding projector $\hat{P}_{1}=\ket{1}_{b}\bra{1}$. In Fig.~\ref{fig:3.3} we plot the $i_1;i_2$ logarithmic negativity heralded from both a `click' detection as well as a 1-PNR detection.  We also include the heralding probability for reference.  For short interaction times, both detection methods are in agreement owing to the probability of obtaining $>1$ photons in the SH-mode being effectively zero.  As the SH-mode becomes more populated, the `click' probability increases though at the expense of smaller degrees of entanglement formation between modes.  We do point out, however, that this only occurs for experimentally inaccessible interaction times; for realistic interaction times, the heralding probability is closer to $\sim0.1$\%. Further still, this entanglement formation is a low-average photon number effect wherein the entanglement scales poorly with initial average photon number $\bar{n}_{s}(0)$.  Unlike the previously mentioned CV entanglement swapping schemes \cite{ref:Li}\cite{ref:Kumar}\cite{ref:Jia}\cite{ref:Marshall}, no post-processing of the measurement results needs to be performed for our proposed experimental set-up.  If such an experiment were to be performed, entanglement forms upon either a `click' or 1-PNR detection. 

\section{Appendix C: Physical determination of scaled dimensionless time $\tau$}\label{sec:Appendix_C}
As discussed shortly after equation \eqref{eqn:trilinear_2}, the full state after SHG is $|\Psi(\tau)\rangle=\hat{U}_{T}|\Psi(0)\rangle$, where:
\begin{equation}
\hat{U}_{T}=e^{-\frac{i}{\hbar}\int_{0}^{T}dt'\hat{H}_{I}}.
\end{equation}
Using the formalism in \cite{schneeloch2019introduction} describing the Hamiltonian in $\chi^{(2)}$ nonlinear-optical interactions, we can obtain a more verbose trilinear hamiltonian coupling modes $(s_{1},s_{2})$ to modes $b$:
\begin{widetext}
\begin{equation}
\hat{H}_{I}=2i d_{eff}\sqrt{\frac{\hbar^{3}\omega_{b}\omega_{s_{1}}\omega_{s_{2}}}{2\epsilon_{0}L_{z}^{3}n_{b}^{2}n_{s_{1}}^{2}n_{s_{2}}^{2}}}\Phi(\Delta k_{z})e^{-i\Delta\omega t}\hat{a}_{s_{1}}\hat{a}_{s_{2}}\hat{b}^{\dagger} + h.c.
\end{equation}
where $\Phi(\Delta k_{z})$ is the spatial overlap integral:
\begin{equation}
\Phi(\Delta k_{z})\equiv\int d^{3}r\;\left(\bar{\chi}_{eff}^{(2)}(\vec{r}) g_{s1}(x,y) g_{s2}(x,y)g_{b}^{*}(x,y)  e^{-i\Delta k_{z} x}\right)
\end{equation}
\end{widetext}
From this, we see that much of $\hat{H}_{I}$ is consolidated into the coupling constant $\kappa$
\begin{equation}
\kappa = 2 d_{eff}\sqrt{\frac{\hbar\omega_{b}\omega_{s_{1}}\omega_{s_{2}}}{2\epsilon_{0}L_{z}^{3}n_{b}^{2}n_{s_{1}}^{2}n_{s_{2}}^{2}}}\Phi(\Delta k_{z})e^{-i\Delta\omega t}
\end{equation}
so that
\begin{equation}
\hat{H}_{I}=i \hbar \kappa (\hat{a}_{s_{1}}\hat{a}_{s_{2}}\hat{b}^{\dagger} - \hat{a}_{s_{1}}^{\dagger}\hat{a}_{s_{2}}^{\dagger}\hat{b})
\end{equation}
Where $t$ is effectively the duration over which the pulse of signal light interacts with the nonlinear medium, we have sufficient information to determine the scaled dimensionless time $\tau = \kappa t$.

For completeness, we define the rest of our parameters:
\begin{itemize}
    \item $\bar{\chi}_{eff}^{(2)}(\vec{r})$ is a scaled function for the nonlinearity of the material. It is zero outside the material, and is assumed to be unity inside the material for a constant nonlinearity overall. If the material is periodically poled, then $\bar{\chi}_{eff}^{(2)}$ flips between $1$ and $-1$ within the material as the poling flips. 
    \item $g_{b}(x,y)$ is the transverse mode amplitude of the SHG mode ($b$). It is normalized so that its magnitude square integrated over all transverse space gives unity. $g_{s_{1}}(x,y)$ and $g_{s_{2}}(x,y)$ are similarly defined for signal modes $s_{1}$ and $s_{2}$. We will typically take these to be simple gaussian functions with effective diameter corresponding to the effective diameter of the beam of light incident on or exiting the crystal. 
    \item $\Delta k_{z}$ is the longitudinal momentum mismatch such that $\Delta k_{z}= k_{bz} - k_{s_{1}z} - k_{s_{2}z}$, where these momentum values are dependent on the frequency of the light and the dispersion of the material. Similarly, we have that $\Delta\omega=\omega_{b}-\omega_{s_{1}}=\omega_{s_{2}}$.
    \item $d_{eff}$ is the effective nonlinearity of the material, which is a constant of the material when accounting the orientation of the crystal axes relative to the polarization of the incident and generated light.
    \item $L_{z}$ is the length of the nonlinear medium through which the light is propagating.
    \item $n_{b}$, $n_{s_{1}}$, and $n_{s_{2}}$ are the indices of refraction at the SHG frequency and at the signal mode frequencies, respectively.
    \item For SHG, we have that the the two signal angular frequencies $\omega_{s_{1}}$ and $\omega_{s_{2}}$ are equal to each other, and to half the SHG angular frequency $\omega_{b}$.
\end{itemize}

Because we are performing SHG, we have that $\Delta\omega=0$,  making $\hat{H}_{I}$ constant in time.

Using the assumption that the spatial modes are gaussian with radii $\sigma_{s1}=\sigma_{s2}=\sigma_{b}\sqrt{2}$, and that for SHG $\omega_{s_{1}}=\omega_{s_{2}}=\omega_{b}/2$, we can solve the overlap integral to find:
\begin{equation}
\kappa = \frac{d_{eff}}{2\sigma_{b}}\sqrt{\frac{\hbar\omega_{b}^{3}}{\pi \epsilon_{0}L_{z}n_{b}^{2}n_{s_{1}}^{2}n_{s_{2}}^{2}}}\text{sinc}\left(\frac{\Delta k_{z} L_{z}}{2}\right)
\end{equation}
If we further consider only pairs of frequencies that are nearly perfectly phase-matched, we may approximate the sinc function as unity.

In this simplified case, we can find a typical value \footnote{Typical values used $d_{eff}$ from $1-30$pm/V; used $L_{z}$ from $1-10$mm; used $\lambda_{b}=775$nm; used $n_{b}=n_{s_{1}}=n_{s_{2}}=1.8$; and used $\sigma_{b}$ from $20-200$ microns.} of $\kappa$ ranging from $10^{3}$ to $10^{6}$, which, when multiplied by a pulse duration $t$ of $1$ns produces scaled dimensionless times $\tau$ of the order $10^{-6}$ to $10^{-3}$. 


\bibliography{SHG_main}

\end{document}